\documentclass[11pt, draftclsnofoot, onecolumn]{IEEEtran}

\usepackage{cite}
\usepackage{amsmath,amsthm}
\usepackage{array}
\usepackage{amssymb}
\usepackage{array}
\usepackage{cite}
\usepackage{ifthen}
\usepackage{graphicx,subfigure}
\usepackage{psfrag}
\usepackage{dsfont}
\usepackage{color}
\renewcommand{\baselinestretch}{1.5}

\newcommand{\ub}{\underline{b}}
\newcommand{\ud}{\underline{d}}
\newcommand{\uw}{\underline{w}}
\newcommand{\us}{\underline{s}}
\newcommand{\ubx}{\underline{\boldsymbol{x}}}
\newcommand{\uby}{\underline{\boldsymbol{y}}}
\newcommand{\bx}{{\boldsymbol{x}}}
\newcommand{\by}{{\boldsymbol{y}}}
\newcommand{\hbb}{\underline{\hat{b}}}
\newcommand{\bz}{\boldsymbol{z}}
\newcommand{\ubz}{\underline{\boldsymbol{z}}}
\newcommand{\ul}{\underline{l}}
\newcommand{\ppi}{p_\mathrm{i}}
\newcommand{\ppd}{p_\mathrm{d}}
\newcommand{\ppt}{p_\mathrm{t}}
\newcommand{\ppid}{p_\mathrm{id}}

\begin{document}
\title{Reliable communication over non-binary insertion/deletion channels}
\author{\vspace{0.2in}
        \normalsize Raman~Yazdani,~\IEEEmembership{\normalsize Student~Member,~IEEE,}
        and
        \normalsize Masoud~Ardakani,~\IEEEmembership{\normalsize Senior~Member,~IEEE}%
        \\
        \normalsize{Department of Electrical and Computer Engineering,
        University of Alberta\\
        Edmonton, Alberta, T6G 2V4, Canada\\
        \{yazdani, ardakani\}@ece.ualberta.ca}}
\maketitle

\begin{abstract}
We consider the problem of reliable communication over non-binary insertion/deletion channels where symbols are randomly deleted from or inserted in the transmitted sequence and all symbols are corrupted by additive white Gaussian noise. To this end, we utilize the inherent redundancy achievable in non-binary symbol sets by first expanding the symbol set and then allocating part of the bits associated with each symbol to watermark symbols. The watermark sequence, known at the receiver, is then used by a forward-backward algorithm to provide soft information for an outer code which decodes the transmitted sequence. Through numerical results and discussions, we evaluate the performance of the proposed solution and show that it leads to significant system ability to detect and correct insertios/deletions. We also provide estimates of the maximum achievable information rates of the system, compare them with the available bounds, and construct practical codes capable of approaching these limits.
\end{abstract}


\section{Introduction}
Since the seminal work of Shannon \cite{shannon48}, there have been huge advancements in coding and information theory. The fundamental limits and efficient coding solutions approaching these limits are now known for many communication channels. However, in the vast majority of coding schemes invented, it is assumed that the receiver is perfectly synchronized with the transmitter, i.e., the symbol arrival times are known at the receiver. In most communication systems, however, achieving perfect synchronization is not possible even with the existence of timing recovery systems.

When perfect synchronization does not exist, random symbol insertions and deletions (synchronization errors) occur in the received sequence. This phenomenon poses a great challenge for error correction. Since the positions of the inserted/deleted symbols are unknown at the receiver, even a single uncorrected insertion/deletion can result in a catastrophic burst of errors. Thus, conventional error-correcting codes fail at these situations.

Error-correcting codes designed for dealing with such insertion/deletion (I/D) channels are called {\it synchronization codes}. Synchronization codes have a long history but their design and analysis has proven to be extremely challenging, hence few practical results exist in the literature. {\color{black}Moreover, standard approaches do not lead to finding the optimal codebooks or tight bounds on the capacity of I/D channels and finding their capacity is still an open problem \cite{mitzenmacher09}.}

The first synchronization code was proposed by Sellers in 1962 \cite{sellers62}. He inserted
{\it marker} sequences in the transmitted bitstream to achieve
synchronization. Long markers allowed the decoder to correct
multiple insertion or deletion errors but greatly increased the
overhead. In 1966, using number-theoretic techniques, Levenshtein constructed
binary codes capable of correcting a single insertion or deletion assuming
that the codeword boundaries were known at the decoder
\cite{levenshtein66}. Most subsequent work were inspired by the
number-theoretic methods used by Levenshtein, e.g., see \cite{tenegolts76,helberg93,saowapa00,helberg02}. Unfortunately, these constructions either cannot be generalized to correct multiple synchronization errors without a significant loss in rate, do not scale well for large block lengths, or lack practical and efficient encoding or decoding algorithms.

Some authors also generalized these number-theoretic methods to non-binary alphabets and constructed non-binary synchronization codes \cite{calabi69,tanaka76,tenegolts84,levenshtein92,levenshtein02}. Following \cite{levenshtein92}, \emph{perfect} deletion-correcting codes were studied and constructed using combinatorial approaches \cite{yin01,klein04,wang06}. Most of these codes, however, are constructed using ad hoc techniques and no practical encoding and decoding algorithm is provided. Non-binary low-density parity-check (LDPC) codes decoded by a verification-based decoding algorithm are designed for deletion channels in \cite{mitzenmacher06}. Unfortunately, the decoding complexity of this construction is also far from being practical.

%

The drawback of all the above-mentioned synchronization codes is that they only work under very stringent synchronization and noise restrictions such as working only on deletion channels, or a single synchronization error per block. Coding methods proposed for error-correction
on the I/D channels working under more general conditions are usually based on concatenated coding schemes with two layers of codes, i.e., an inner and an outer code \cite{schulman99,davey01,chen03,ratzer05,buttigieg11}. The inner code identifies the positions of the synchronization errors and the outer code is responsible for correcting the insertions, deletions, and substitution errors as well as misidentified synchronization errors.

In the seminal work of Davey and MacKay \cite{davey01}, a practical concatenated coding method is presented for error-correction on general binary I/D channels. They have called their inner code, a \emph{watermark} code. The main idea is to provide a carrier signal or watermark for the outer code. The synchronization errors are inferred by the outer code via identifying discontinuities in the carrier signal. One of the advantages of watermark codes is that the decoder does not need to know the block boundaries of the received sequence. However, due to the use of a sparsifier, rate loss is significant. The watermark is substituted by fixed and pseudo-random markers in \cite{ratzer05} and is shown that it allows better rates but is only able to outperform the watermark codes at low synchronization error rates. {\color{black}Also, it has recently been shown that the performance of watermark codes can be improved by using symbol-level decoding instead of bit-level decoding \cite{briffa10,wang11}}.


In this work, we consider the problem of \textcolor{black}{devising an efficient coding method for} reliable communication over non-binary I/D channels. On these channels, synchronization errors occur at the symbol level, i.e., symbols are randomly inserted in and deleted from the received sequence. We also assume that all symbols are corrupted by additive white Gaussian noise (AWGN). {\color{black}The use of this channel model is motivated by the fact that at the receiver the received continuous waveform is first sampled at certain time instances to produce the discrete symbol sequence required by the decoder. If the symbol arrival times are not perfectly known at the receiver, i.e., there is timing mismatch, some of the transmitted symbols are not sampled at all (symbol deletions) or sampled multiple times (symbol insertions) \cite{barry_tutorial}. As a result, this channel model can be used to represent non-binary communications over the AWGN channel suffering from timing mismatch. Most communication systems use non-binary signalling, where synchronization errors can result in insertion/deletion at the symbol level.}

For the proposed channel model, we utilize the inherent redundancy that can be achieved in non-binary symbol sets by first expanding the symbol set and then allocating part of the bits associated with each symbol to watermark symbols. As a result, not all the available bits in the signal constellation are used for the transmission of information bits. In its simplest form, our solution can be viewed as a communication system using two different signal sets. The system switches between these two signal sets according to a binary watermark sequence. Since the watermark sequence is known both at the transmitter and the receiver, probabilistic decoding can be used to infer the insertions and deletions that occurred and to remove the effect of additive noise. In particular, the system is modeled by a hidden Markov model (HMM) \cite{rabiner86} and the forward-backward algorithm \cite{bahl74} is used for decoding.

Our proposed scheme resembles trellis coded modulation (TCM) \cite{ungerboeck82}. The main idea in both methods is to add redundancy by expanding the symbol set and limit the symbol transitions in a controlled manner. The proposed method is also closely related to the watermark codes of \cite{davey01}. In both methods, decoding is done by the aid of a watermark sequence which both the transmitter and receiver agree on. The difference is that the extra degree of freedom in non-binary sets allows us to separate information from the watermark.

Our proposed solution leads to significant system ability to detect and correct synchronization errors. For example, a rate $1/4$ binary outer code is capable of correcting about $2,900$ insertion/deletion errors per block of $10,012$ symbols \textcolor{black}{even when block boundaries are unknown at the receiver}.

This paper is organized as follows. In Sections \ref{Sec:model_approach} and \ref{sec:system_model}, we state our proposed approach and describe the system model. Section \ref{Sec:error_analysis} demonstrates the capabilities of the proposed solution by providing numerical results and discussions. Section~\ref{sec:increasing_achievable_rates} describes ways to increase the achievable information rates on the channel and Section~\ref{sec:complexity} analyzes the system in terms of complexity, and practical considerations. Finally, Section \ref{Sec:conclusion} concludes the paper.

\section{Channel model and the proposed approach} \label{Sec:model_approach}
Throughout this paper, scalar quantities are shown by lower case symbols, complex quantities by boldface letters, and vectors by underlined symbols.
\subsection{Channel model}
The channel model we consider in this work is a non-binary I/D channel with AWGN where insertions and deletions occur at the symbol level. Similar to \cite{davey01}, it is assumed that the symbols from the input sequence $\ubx$ first enter a queue before being transmitted. Then at each channel use, either a random symbol is inserted in the symbol sequence $\ubx'$ with probability $p_\mathrm{i}$, the next queued symbol is deleted with probability $p_\mathrm{d}$, or the next queued symbol is transmitted (put as the next symbol in $\ubx'$) with probability $p_\mathrm{t}=1-p_\mathrm{d}-p_\mathrm{i}$. For computational purposes, we assume that the maximum number of insertions which can occur at each channel use is $I$. The resulting symbol sequence $\ubx'$ is finally affected by an i.i.d. sequence of AWGN $\ubz$ where $\bz\sim\mathcal{CN}(0,2\sigma^2)$ and $\uby=\ubx'+\ubz$ is received at the receiver side.

{\color{black}Note that in this paper, to show the capabilities of the proposed method, we consider totally random and independent symbol insertions/deletions. When symbol insertions are resulted from imperfect synchronization, insertions or deletions tend to be correlated. These cases lead to easier identification of insertions/deletions at the receiver compared to random independent insertions/deletions which we consider here.}

\subsection{Proposed approach}

Now, consider a communication system working on this channel by employing an $M$-ary signalling (e.g., $M$-ary phase-shift keying (PSK)). We call this the \emph{base} system. Motivated by the idea of watermark codes \cite{davey01}, we are interested in embedding a watermark in the transmitted sequence. The watermark, being known at the receiver, allows the decoder to deduce the insertions and deletions and to recover the transmitted sequence.

{\color{black}The watermark can be embedded in the transmitted sequence in many ways. One way of doing this is to add the watermark to the information sequence and treat the information sequence as additive noise at the receiver. This is a direct extension of the binary watermark codes of \cite{davey01} to non-binary signalling. In particular, the additive watermark $\uw$ can be defined as a sequence of $M$-ary symbols drawn from the base system constellation. The binary information sequence is first passed through a sparsifier; every $k$ bits of the information sequence is converted to an $n$-tuple of $M$-ary symbols. The rate of the sparsifier is then given by $r_\mathrm{s}=k/n$ where $0<r_\mathrm{s}<m$ and $M=2^m$. The average density of the sparsifier $f$ is defined as the average Hamming distance of the $n$-tuples divided by $n$. The mapping used in the sparsifier is chosen as to minimize $f$.

By defining addition as shifting over the constellation symbols, the watermark sequence could be added to the sparsified messages (denote it by $\us$) and $\uw\oplus\us$ be sent over the channel. At the receiver, similar to \cite{davey01}, an inner decoder which knows the watermark sequence, uses the received sequence to deduce the insertions/deletions and provides soft information for an outer code.

The main drawback of this method is that the decoder is not able to distinguish between additive noise and the information symbols. This is because the information is embedded into the watermark by adding $\us$ to $\uw$. Sequence $\us$ contains both zeros and non-zero symbols. Non-zero symbols shift the watermark symbols over the constellation, similar to what additive noise does. This greatly degrades the performance of the decoder. To improve the decoding performance, $\us$ should contain as many zeros as possible, i.e., be as sparse as possible, which is equivalent to having a small $f$. A small $f$ is achieved by decreasing $r_\mathrm{s}$ which in turn decreases the achievable rates on the channel directly. Also, notice that even in the absence of additive noise, the decoder is still fooled by the shifts occurred over $\uw$ and thus misidentifies some of the insertions/deletions.}

\smallskip

To aid error recovery at the receiver, we are interested in an embedding method which makes the watermark as distinguishable as possible from the information sequence. This necessitates using some \emph{extra} resources (other than those used to transmit the information sequence) for transmitting the watermark sequence. These extra resources can be provided by enlarging the signal set. The extra available bits per transmission can then be used to transmit the watermark. After embedding the watermark, we refer to the system as the \emph{watermarked} system.

In this work, we are mostly interested in binary watermark sequences. As a result, to accommodate the watermark bits in each symbol, we expand the signal set size $M$ by the factor of $2$, giving rise to a $2M$-ary signalling scheme. For example, if the base system uses 4-PSK, in the watermarked system we use 8-PSK modulation. To provide fair comparison, we put the symbol rate, information bits per symbol \textcolor{black}{(denoted by $r_\mathrm{c}$)}, and average energy of the signal constellation of the watermarked system equal to those of the base system. As a result, the spectral efficiency and the total transmitted power of the watermarked system are equal to those of the base system. In other words, no bit rate, bandwidth, or power is sacrificed as a result of embedding the watermark.

{\color{black}Notice that $r_\mathrm{c}=m$ where $M=2^m$ for the base system and also the watermarked system when a binary watermark sequence is used for each transmitted symbol. This is because in an $M$-ary base system all the $m$ available bits are dedicated to information bits. Also, in each symbol of the $2M$-ary watermarked system (with $m+1$ available bits) $m$ bits are assigned to information bits. Later, we will see that sometimes it is more efficient to use non-binary watermark sequences or to assign less than one bit per symbol on average to the watermark giving rise to $0<r_\mathrm{c}<m+1$. These cases will be investigated in Section \ref{sec:increasing_achievable_rates}.}

Expanding the signal set while fixing the average energy of the constellation leads to reduction in the minimum distance of the constellation. Nevertheless, we show that by using the mapping described in Section \ref{Sec:modulator}, the minimum distance $d_\mathrm{min}$ between symbols corresponding to the same watermark value does not necessarily reduce. In fact in some cases, e.g., in PSK modulation, the minimum distance does not change compared to the base system. Thus, the noise immunity\footnote{Here, the noise immunity is measured in the absence of synchronization errors under the assumption of minimum distance decoding. As a result, the minimum distance between the signal constellation points can be used as the noise immunity measure.} of the system does not change after adding the watermark.

\section{System model}\label{sec:system_model}
\begin{figure}
\centering
\psfrag{b}{$\ub$}
\psfrag{d}{$\ud$}
\psfrag{w1}{$\uw$}
\psfrag{w2}{$\uw$}
\psfrag{x}{$\ubx$}
\psfrag{y}{$\uby$}
\psfrag{l}{$\ul$}
\psfrag{bhat}{$\hbb$}
\psfrag{z}{$\ubz$}
\includegraphics[width=.7\columnwidth]{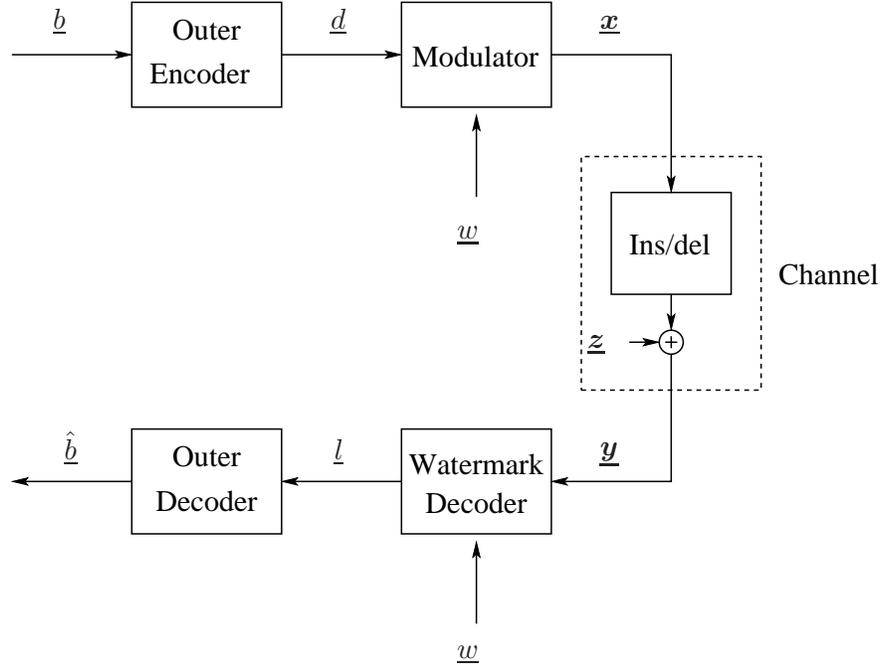}\\
\caption{The proposed system model.}\label{fig:system_model}
\end{figure}

The proposed system model is shown in Fig.~\ref{fig:system_model}. First, the binary information sequence $\ub$ is encoded by the outer code producing the coded binary sequence $\ud$ which is then broken into $m$-bit subsequences. The modulator then combines the binary watermark $w$ and the $m$-bit subsequences by a one-to-one mapping $\mu:\{0,1\}^{m+1}\rightarrow\mathcal{X}$ where $\mathcal{X}$ is the signal set of size $|\mathcal{X}|=2M=2^{m+1}$. Then $\ubx$ is sent over the channel. The received sequence $\uby$ is first decoded by the watermark decoder which provides soft information for the outer decoder in terms of log-likelihood ratios (LLRs). The LLR sequence $\ul$ is then utilized to decode the information sequence $\hat{\ub}$.

\subsection{Modulator} \label{Sec:modulator}
The modulator plays a key role in the proposed system. It allows embedding encoded data and watermark bits while ensuring a good minimum distance. The most important part of designing the modulator is to choose an appropriate mapping $\mu$. By viewing $\mu$ as $\{0,1\}\times\{0,1\}^m\rightarrow\mathcal{X}$, we first divide $\mathcal{X}$ into two disjoint subsets $\mathcal{X}^0$ and $\mathcal{X}^1$ each having $M$ signal points corresponding to watermark bit $w=0$ and $w=1$, respectively. Thus, in the label of each signal point, one bit (can be any of the $m+1$ bits) is dedicated to the watermark bit and the other $m$ bits correspond to the $m$-bit subsequences of $\ud$.
Formally we have
\begin{equation*}\label{eq:subsets}
    \mathcal{X}^{w}=\left\{\bx|\bx\in\mathcal{X};\ell_\mathrm{w}(\bx)=w\right\},\quad\mathrm{for}\quad w=0,1,
\end{equation*}
where $\ell_{\mathrm{w}}(\bx)$ denotes the value of the bit in the label of $\bx$ dedicated to the watermark. We also define $\ell^j(\bx)$ for $j=1,2,\dots,m$ as the $j$-th non-watermark bit of the label of $\bx$.

\begin{figure}
\centering
    \subfigure[Base system]
    {
        \psfrag{$d_{min}$}{$\scriptstyle d_{\mathrm{min}}$}
      \label{fig:4_PSK}{\includegraphics[width=.45\columnwidth]{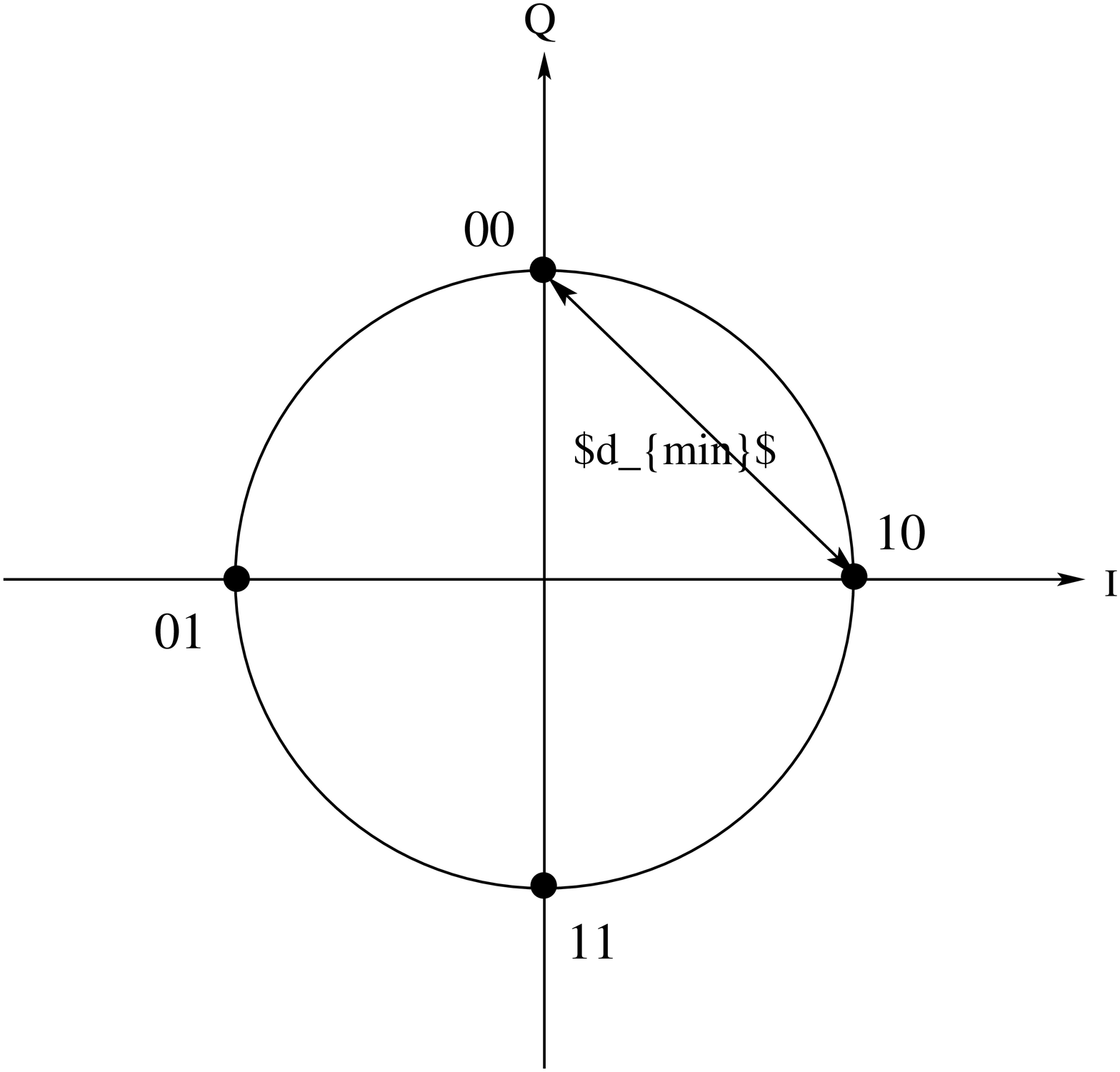}}
    }
    \hfill
    \subfigure[Watermarked system]
    {
        \psfrag{$d_{min}$}{$\scriptstyle d_{\mathrm{min}}$}
      \label{fig:8_PSK}{\includegraphics[width=.45\columnwidth]{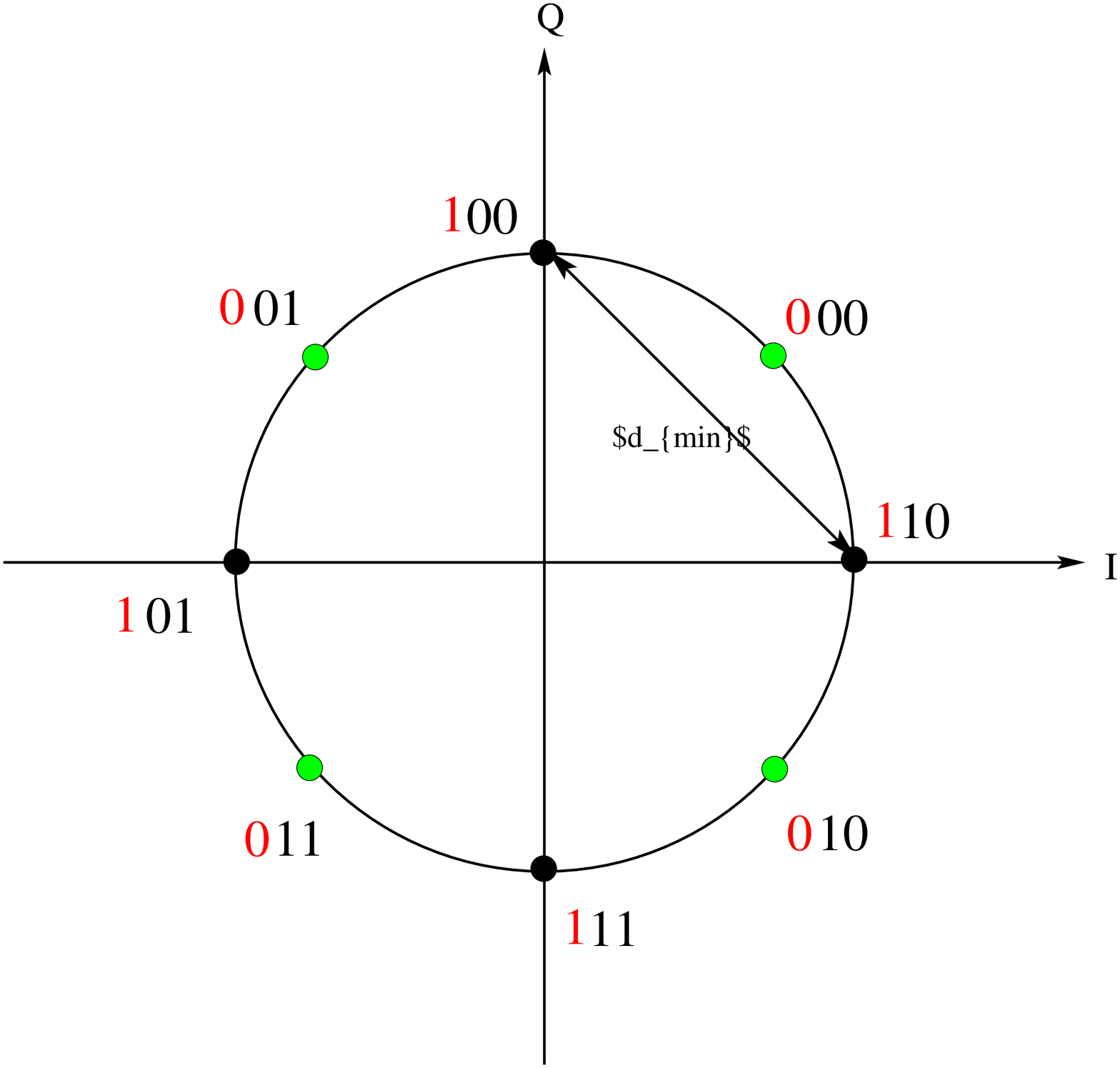}}
    }
  \caption{Signal constellations and their labeling for the base system ($4$-PSK) and the watermarked system ($8$-PSK). The leftmost bit in the label of the watermarked system corresponds to the watermark bit. For both constellation $d_\mathrm{min}=\sqrt{2}$.}
  \label{fig:4_8_PSK}
\end{figure}

\begin{figure}
\centering
    \subfigure[Base system]
    {
    \psfrag{dmin}{$\scriptstyle d_{\mathrm{min}}$}
      \label{fig:16_QAM}{\includegraphics[width=.45\columnwidth]{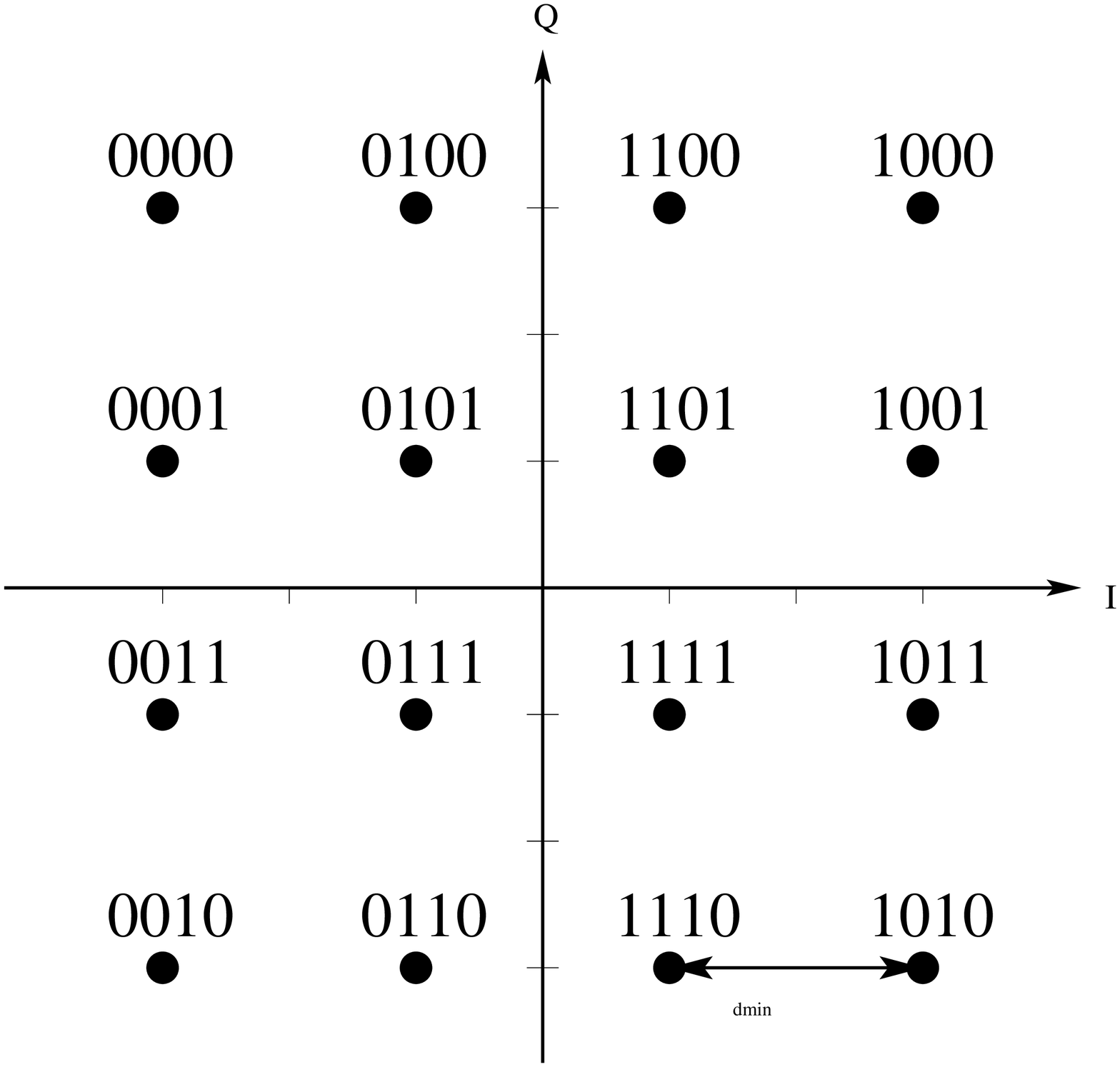}}
    }
    \hfill
    \subfigure[Watermarked system]
    {
    \psfrag{dmin}{$\scriptstyle d_{\mathrm{min}}$}
      \label{fig:32_QAM}{\includegraphics[width=.45\columnwidth]{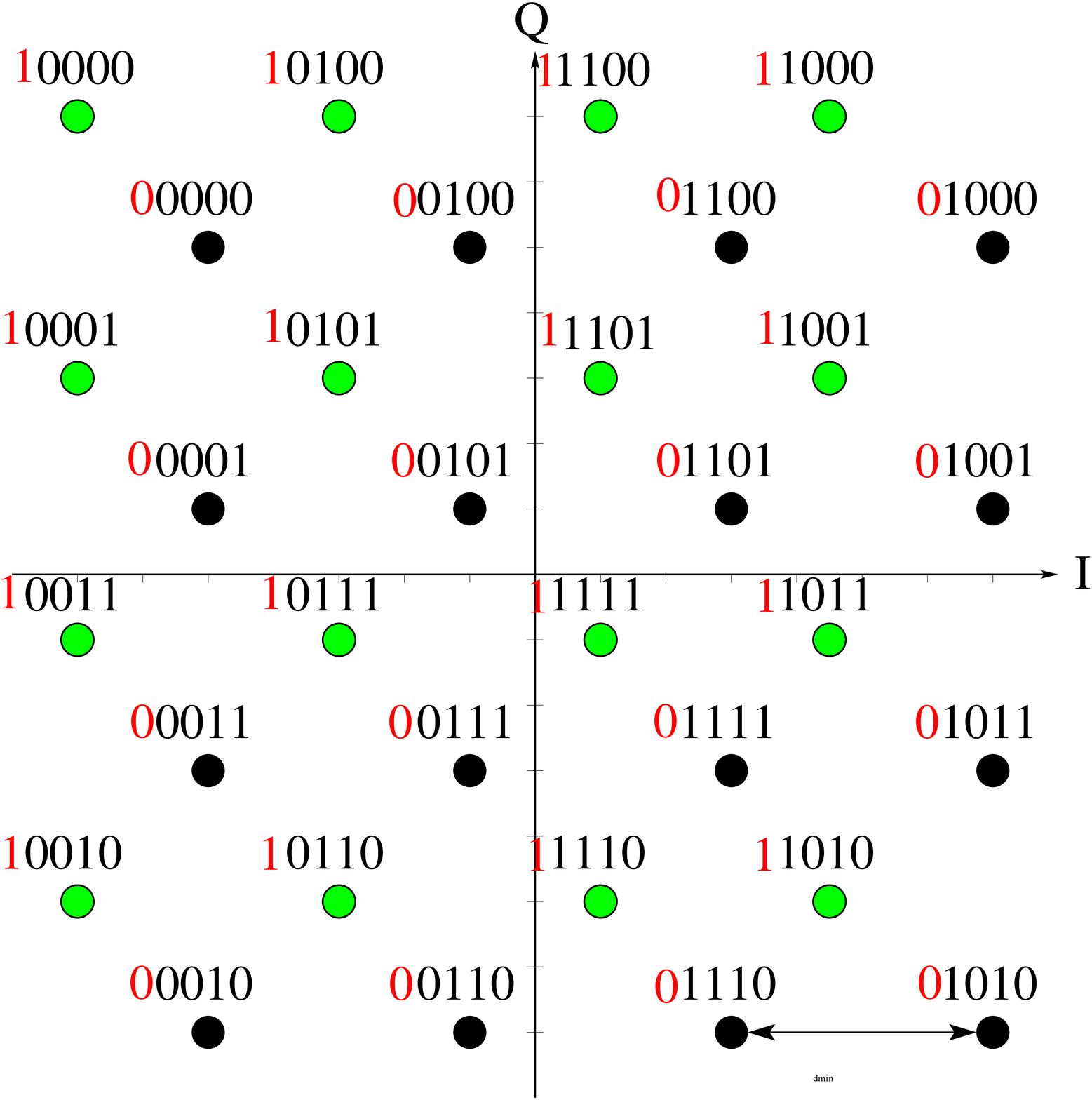}}
    }
  \caption{Signal constellations and their labeling for the base system ($16$-QAM) and the watermarked system ($32$-AM/PM). The leftmost bit in the label of the watermarked system corresponds to the watermark bit. Both constellation have unit average energy. Thus, $d_\mathrm{min}=2/\sqrt{10}=0.633$ for the base system and $d_\mathrm{min}=4/\sqrt{42}=0.617$ for the watermarked system.}
  \label{fig:16_32_QAM}
\end{figure}

Now the question is how to choose the labeling. To maximize the noise immunity of the system, and since the watermark sequence is known at the receiver, we maximize the minimum distance between the signal points in each of $\mathcal{X}^0$ and $\mathcal{X}^1$. To do this, first we do a one-level set partitioning \cite{ungerboeck82}, i.e., we divide $\mathcal{X}$ into two subsets with the largest minimum distance between the points in each subset. These subsets are named $\mathcal{X}^0$ and $\mathcal{X}^1$ and the watermark bit of the label is assigned accordingly.
Next, by a Gray mapping \cite{caire98} of the signals in each of $\mathcal{X}^0$ and $\mathcal{X}^1$, the non-watermark bits of the label are assigned. This process is illustrated for two different signal constellations in Figs.~\ref{fig:4_8_PSK} and \ref{fig:16_32_QAM}. The Gray mapping ensures the least bit error rate in each subset \cite{caire98}.

The minimum distance of the constellation is now defined as
\begin{equation*}
    d_\mathrm{min} = \min_{w}\min_{\{\bx_i,\bx_j\}\subset\mathcal{X}^w,\bx_i\neq\bx_j}||\bx_i-\bx_j||.
\end{equation*}
Notice that by assuming signal constellations of fixed energy, going from $M$-PSK in the base system to $2M$-PSK in the watermarked system does not change $d_\mathrm{min}$ (see Fig.~\ref{fig:4_8_PSK}). For the QAM, as illustrated in Fig.~\ref{fig:16_32_QAM}, $d_\mathrm{min}$ does change because of energy adjustments but always stays very close to that of the original constellation. For example in Fig.~\ref{fig:16_32_QAM}, $d_\mathrm{min}$ is reduced by only $2.4\%$.

A definition which proves useful in the next sections is
\begin{equation*}\label{eq:labeling}
    \mathcal{X}_j(w_i,d_{i,j})=\left\{\bx|\bx\in\mathcal{X};\ell_\mathrm{w}(\bx)=w_i,\ell^j(\bx)=d_{i,j}\right\},
\end{equation*}
where $i$ denotes the index of both the watermark bit and the $m$-bit subsequences of $\ud$ and $d_{i,j}=d_{(i-1)m+j}$ denotes the $j$-th bit of the $i$-th subsequence. Considering a watermark sequence of length $N$ and an encoded data sequence of length $mN$, then $i=1,2,\dots,N$. Thus, $\mathcal{X}_j(u,v)$ refers to the subset of $\mathcal{X}$ where the watermark bit $w_i$ is equal to $u$ and the $j$-th data bit in the $i$-th subsequence, i.e., $d_{i,j}$, is equal to $v$. The size of this subset is $M/2$.

\subsection{Watermark decoder}\label{sec:watermark_decoder}
The goal of the watermark decoder is to produce LLRs for the outer decoder given $\uw$ and the received sequence $\uby$. As in \cite{davey01}, by ignoring the correlations in $\ud$, we can use an HMM to model the received sequence and then use the forward-backward algorithm \cite{rabiner86} to calculate posterior probabilities or LLRs for the outer decoder. Notice that due to the nature of the channel which introduces insertions and deletions, there will be a synchronization \emph{drift} between $\ubx$ and $\uby$. The synchronization drift at position $i$, i.e., $t_i$ is defined as the (number of insertions) $-$ (number of deletions) occurred in the signal stream until the $i$th symbol, i.e., $\bx_i$, is ready for transmission\footnote{This means that if $\bx_{i-1}$ is not deleted by the channel it is received as $\by_{i-1+t_i}$.}. The drifts ${\{t_i\}}_{i=1}^N$, form the hidden states of the HMM. Each state $t_i$ takes values from
\begin{equation}
\mathbf{T}=\{\dots,-2,-1,0,1,2,\dots\}.
\end{equation}
Thus, $t_i$ performs a random walk on $\mathbf{T}$ whose mean and variance depend on $\ppi$ and $\ppd$. To reduce the decoding complexity, as in \cite{davey01}, we limit the drift to $|t_i|\leq t_\mathrm{max}$ where $t_\mathrm{max}$ is usually chosen large enough such that it accommodates all likely drifts with high probability. For example, when $\ppi=\ppd$, $t_\mathrm{max}$ is chosen several times larger than $\sqrt{N\ppd/(1-\ppd)}$ which represents the standard deviation of the drifts over a block of size $N$.

To further characterize the HMM \cite{rabiner86}, we need the state transition probabilities, i.e., $P_{ab}=P(t_{i+1}=b|t_i=a)$. Each symbol $\bx_i$ entering the channel can produce any number of symbols between $0$ and $I+1$ at the channel output. As a result, if $t_i=a$, then $t_{i+1}\in\{a-1,\dots,a+I\}$. Notice that the transition from $t_i=a$ to $t_{i+1}=b$ can occur in two ways. One is when $\bx_i$ is deleted by the channel and $(b-a+1)$ symbols are inserted by the channel. The other one is when $\bx_i$ is transmitted and $(b-a)$ symbols are inserted by the channel. In either case, $(b-a+1)$ symbols are produced at the channel output. As a result, the state transition probabilities are given by
\begin{equation}\label{eq:state_transition}
P_{ab}=\left\{\begin{array}{ll}
\ppd & b=a-1 \\
\alpha_I\ppi\ppd+\ppt & b=a \\
\alpha_I(\ppi^{b-a+1}\ppd+\ppi^{b-a}\ppt) & a<b<a+I \\
\alpha_I\ppi^I\ppt & b=a+I \\
0 & \mathrm{otherwise}.
\end{array}
\right.
\end{equation}
where $\alpha_I=1/(1-\ppi^I)$ is a constant normalizing the effects of maximum insertion length $I$ and ensuring that the sum of probabilities is $1$.

We also need to calculate the conditional probability of producing the observation sequence $\widetilde{\uby}=\{\by_{i+a},\dots,\by_{i+b}\}$ given the transition from $t_i=a$ to $t_{i+1}=b$. As stated, this transition can occur in two ways. Thus,
\begin{align}
\nonumber Q_{ab}^i(\widetilde{\uby})&=P(\widetilde{\uby}|t_i=a,t_{i+1}=b,w_i,\mathcal{H})\\
&=\left(\alpha_I\ppi^{b-a+1}\ppd\prod_{k=i+a}^{i+b}\gamma_k+
\alpha_I\ppi^{b-a}\ppt\beta_{i+b}\prod_{k=i+a}^{i+b-1}\gamma_k\right)/P_{ab},
\end{align}
where $\mathcal{H}$ denotes set of parameters of the HMM, i.e., $\mathcal{H}=\{[P_{ab}],\mathbf{T}\}$, $\gamma_k$ is the probability of receiving $\by_k$ given that $\by_k$ is an inserted symbol, and $\beta_{i+b}$ is the probability of receiving $\by_{i+b}$ assuming that it is the result of transmitting $\bx_i\in\mathcal{X}^{w_i}$. Formally, we have
\begin{equation}
\gamma_k=\frac{1}{2M}\sum_{\bx\in\mathcal{X}}\frac{1}{2\pi\sigma^2}
\exp{\left(-\frac{|\by_k-\bx|^2}{2\sigma^2}\right)},
\end{equation}
and
\begin{align}
\nonumber\beta_{i+b}=P(\by_{i+b}|t_i=a,t_{i+1}=b,w_i,\mathcal{H})
=\frac{1}{M}\sum_{\bx\in\mathcal{X}^{w_i}}\frac{1}{2\pi\sigma^2}
\exp{\left(-\frac{|\by_k-\bx|^2}{2\sigma^2}\right)}.
\end{align}

Now that the HMM is defined, we use the forward-backward algorithm to calculate LLRs. By ignoring the correlations between the bits of $\ud$ \textcolor{black}{and assuming $P(d_{i,j}=0) = P(d_{i,j}=1)$}, the bit by bit LLR is calculated as
\begin{align}\label{eq:LLR}
    l_{i,j}&=\log\frac{P(d_{i,j}=0|\uby,\uw,\mathcal{H})}{P(d_{i,j}=1|\uby,\uw,\mathcal{H})}=
    \log\frac{P(\uby|d_{i,j}=0,\uw,\mathcal{H})}{P(\uby|d_{i,j}=1,\uw,\mathcal{H})}
    =\log\frac{\sum_{\bx_i\in\mathcal{X}_j(w_i,0)}P(\uby|\bx_i,\uw,\mathcal{H})}
    {\sum_{\bx_i\in\mathcal{X}_j(w_i,1)}P(\uby|\bx_i,\uw,\mathcal{H})}.
\end{align}
By using the forward-backward algorithm, the posterior probabilities are found by \cite{rabiner86,davey01}
\begin{equation}\label{eq:prob_after_LLR}
P(\uby|\bx_i,\uw,\mathcal{H})=\sum_{a,b}F_i(a)\acute{Q}^i_{ab}(\widetilde{\by}|\bx_i)B_{i+1}(b)
\end{equation}
where the forward quantity is defined as
\begin{equation}
F_i(a)=P(\by_1,\dots,\by_{i-1+a},t_i=a|\uw,\mathcal{H}),
\end{equation}
the backward quantity as
\begin{equation}
B_i(b)=P(\by_{i+b},\dots|t_i=b,\uw,\mathcal{H}),
\end{equation}
and
\begin{align}
\nonumber\acute{Q}^i_{ab}(\widetilde{\by}|\bx_i)&=P(\widetilde{\by},t_{i+1}=b|t_i=a,\bx_i,\mathcal{H})\\
&=\alpha_I\ppi^{b-a+1}\ppd\prod_{k=i+a}^{i+b}\gamma_k+
\alpha_I\ppi^{b-a}\ppt\acute{\beta}_{i+b}\prod_{k=i+a}^{i+b-1}\gamma_k,
\end{align}
where
\begin{align}
\nonumber\acute{\beta}_{i+b}=P(\by_{i+b}|t_i=a,t_{i+1}=b,\bx_i,\mathcal{H})
=\frac{1}{2\pi\sigma^2}\exp{\left(-\frac{|\by_{i+b}-\bx_i|^2}{2\sigma^2}\right)}.
\end{align}
The forward and backward quantities are recursively computed by the forward pass
\begin{equation}
F_i(a)=\sum_{c\in \{a-I,\dots,a+1\}}F_{i-1}(c)P_{ca}Q^{i-1}_{ca}(\by_{i-1+c},\dots,\by_{i-1+a}),
\end{equation}
and the backward pass
\begin{equation}
B_i(b)=\sum_{c\in \{b-1,\dots,b+I\}}P_{bc}Q^i_{bc}(\by_{i+b},\dots,\by_{i+c})B_{i+1}(c).
\end{equation}

{\color{black}If the block boundaries are not known at the decoder, we can use the sliding window decoding technique used in \cite{davey01}. Assuming a continuous stream of transmitted blocks and received symbols, the forward-backward algorithm is used to infer the block boundaries. Then the decoding window is anchored at the most likely start of the next block and next block is decoded. Most of the results of this paper are shown using this sliding window decoding technique. We will briefly explain the methodology in Section~\ref{Sec:error_rates}. For the first transmitted block, we assume that the initial drift is zero. Thus, we use}
\begin{equation}\label{eq:initial_condition_forward}
F_1(a)=\left\{\begin{array}{ll}
1 & a=0 \\
0 & \mathrm{otherwise}.
\end{array}
\right.
\end{equation}

{\color{black}It is also possible to insert some markers which specify the block boundaries into the transmitted sequence. By dedicating all the $m+1$ bits in the symbols at the boundaries to markers, they can be easily detected at the receiver. In this case, the block boundaries can be inferred by detecting the markers. Notice that as the block length becomes larger, recognizing the block boundaries requires less overhead and becomes more efficient. To see this, first define the marker rate as the number of marker symbols in each block divided by $N$. Given a fixed marker rate, the number of marker symbols is increased as $N$ grows. Increasing the number of marker symbols leads to a better boundary detection at the decoder for a fixed $\ppd$ and $\ppi$ because the probability of misidentifying larger block of markers is decreased. Thus, for large block lengths one can assume that the block boundaries are known at the decoder and use the following initial conditions for the backward pass of each block:
\begin{equation}\label{eq:initial_condition_backward}
B_N(b)=P_{bc}Q^N_{bc}(\by_{N+b},\dots)
\end{equation}
where $c=t_{N+1}$ is the final drift at the end of the block.}

%
%


\smallskip
In the next stage of decoding, the LLRs, calculated by inserting (\ref{eq:prob_after_LLR}) in (\ref{eq:LLR}), are passed to the outer decoder.

\subsection{Outer code}
The outer code can almost be any binary error correcting code. Due to the exemplary performance of LDPC codes on many communication channels and their flexible structure, we choose LDPC codes in this paper.

At the transmitter, a binary LDPC code of rate $R$ is used to encode the binary information sequence $\ub$ of length $mNR$ producing the binary coded sequence $\ud$ of length $mN$. At the receiver, the $mN$ bit LLRs of (\ref{eq:LLR}) are used to recover $\ub$.

\section{Error rates and fundamental limits} \label{Sec:error_analysis}
In this section, we demonstrate the capabilities of the proposed solution through examples and discussions. In particular, we evaluate the watermarked system by providing bit and word error rates (BER and WER), maximum achievable transmission rates, and comparing them with two benchmark systems. We demonstrate our results using the following two examples.

\textbf{Example~1:} Consider a base system with $4$-PSK modulation depicted in Fig.~\ref{fig:4_PSK} which gives rise to a watermarked system with $8$-PSK modulation. The labeling $\mu$ is chosen based on the method described in Section \ref{Sec:modulator}. The constellation and labeling are depicted in Fig.~\ref{fig:8_PSK}.

\textbf{Example~2:} In this example, we consider a $16$-QAM base system and a $32$-QAM watermarked system. Notice that different constellations can be considered for $32$-ary modulation. We consider a $32$-AM/PM constellation whose $d_\mathrm{min}$ is very close to the base system (they differ by only $2.4\%$). The constellations and their labelings are depicted in Fig.~\ref{fig:16_32_QAM}.

\subsection{Error rates} \label{Sec:error_rates}

First, consider Example~1. We use a $(3,6)$-regular LDPC code ($R=0.5$) of length $20,024$ constructed by the progressive edge growth (PEG) algorithm \cite{Hu05} as the outer code. The LDPC code is decoded by the sum-product algorithm \cite{richardson01design} allowing a maximum of $400$ iterations. The watermark sequence $\uw$ is chosen to be a pseudo-random binary sequence. Since $m=2$, the block length is $N=10,012$. The maximum insertion length is chosen as $I=5$, the channel insertion and deletion probabilities are assumed equal, i.e., $\ppi=\ppd=\ppid$, {\color{black}and $t_\mathrm{max}=5\sqrt{N\ppid/(1-\ppid)}$.}

{\color{black} A continuous stream of blocks of $\ub$, $\ud$, and $\ubx$ is generated and sent over the channel. A continuous sequence of $\uby$ is then received at the decoder. We assume that block boundaries are not known at the receiver. Thus, $\uby$ is decoded by the forward-backward algorithm using a sliding window decoding technique \cite{davey01}. For the first block, we assume that the receiver knows the starting position, i.e, we use (\ref{eq:initial_condition_forward}) to initialize the forward pass. For subsequent blocks, the watermark decoder is responsible to infer the boundaries and calculate LLRs for the outer code. This is done by first running the forward pass several multiples of $t_\mathrm{max}$ (here, six) beyond the expected position of the block boundary and initializing the backward pass from these last calculated forward quantities. Then the most likely drift at the end of each block is found as $\hat t_{N+1} = \arg\max_aF_{N+1}(a)B_{N+1}(a)$ and is used to slide the decoding window to the most likely start of the next block.

Occasionally, the watermark decoder makes errors in identifying the block boundaries. If these errors accumulate, synchronization is lost and successive blocks fail to be successfully decoded. To protect against such gross synchronization loss, we use the re-synchronization technique of \cite{davey01} whose details are omitted in the interest of space.}

\begin{figure}
\centering
\includegraphics[width=.99\columnwidth]{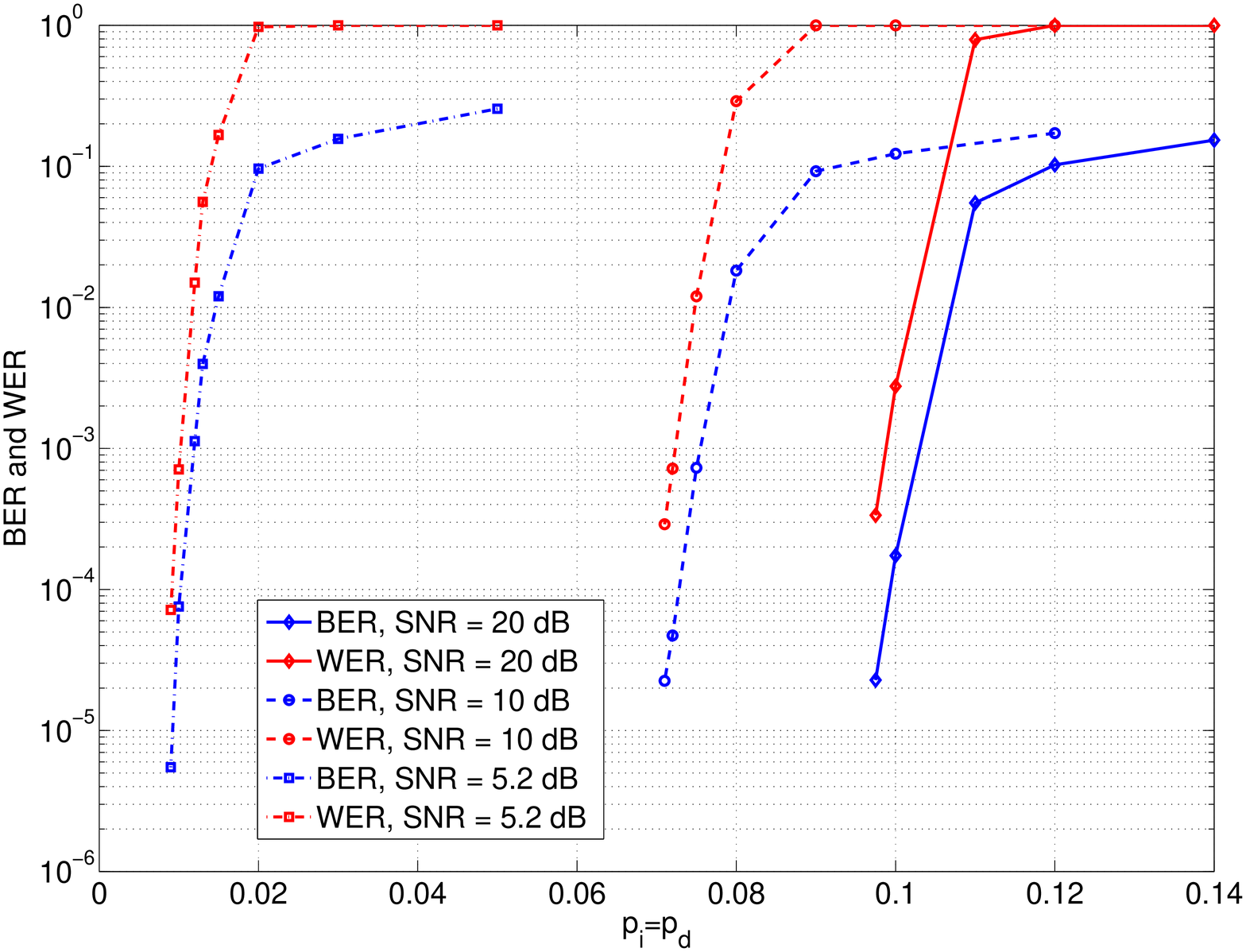}\\
\caption{BER and WER of the $8$-PSK watermarked system employing a (3,6)-regular LDPC code of length $20,024$ versus $\ppid$ at fixed SNRs.}\label{fig:BER_vs_pd}
\end{figure}

\begin{figure}
\centering
\includegraphics[width=.99\columnwidth]{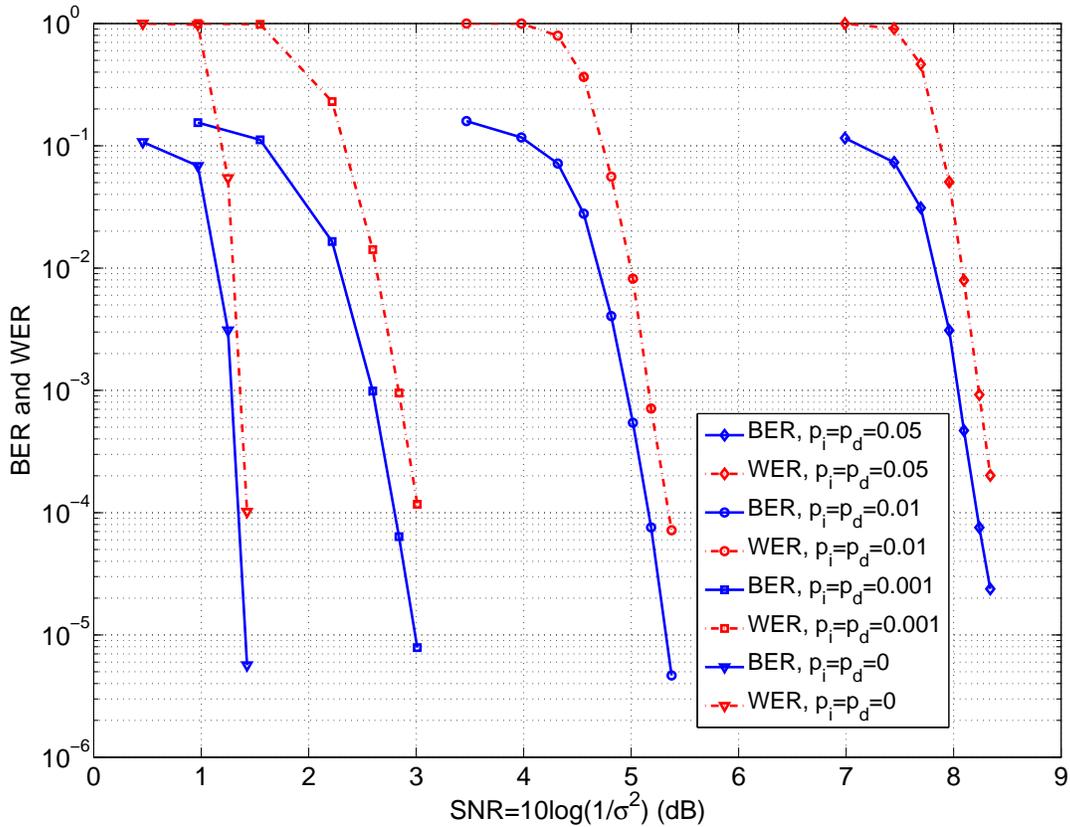}\\
\caption{BER and WER of the $8$-PSK watermarked system employing a (3,6)-regular LDPC code of length $20,024$ versus SNR at fixed values of $\ppid$.}\label{fig:BER_3_6_vs_SNR}
\end{figure}

We simulate the system under different SNRs and insertion/deletion probabilities. The BER and WER of the system are plotted in Fig.~\ref{fig:BER_vs_pd} versus different values of $\ppid$ under fixed SNRs. For example, at SNR$=\!10$~dB, the system is able to recover on average {\color{black}$1,400$} symbol insertions/deletions per block of $10,012$ symbols with an average BER less than $10^{-5}$. This increases to recovering about {\color{black}$1,920$} insertions/deletions at SNR$=\!20$~dB. Fig.~\ref{fig:BER_3_6_vs_SNR} shows the performance of the system versus SNR under fixed $\ppid$.

We also simulate the system under a $(3,4)$-regular LDPC code ($R=0.25$) of length $20,024$ with the same parameters. The system is now capable of recovering on average {\color{black}$2,700$} insertions/deletions per block of $10,012$ symbols at SNR$=\!20$~dB with an average BER$<10^{-5}$. This increases to {\color{black}$2,900$} insertions/deletions using an optimized irregular LDPC code of the same rate and length with degree distributions reported in Table~\ref{Table} (Code~1).
We will briefly describe the LDPC optimization method in the next section.

\begin{figure}
\centering
\includegraphics[width=.75\columnwidth]{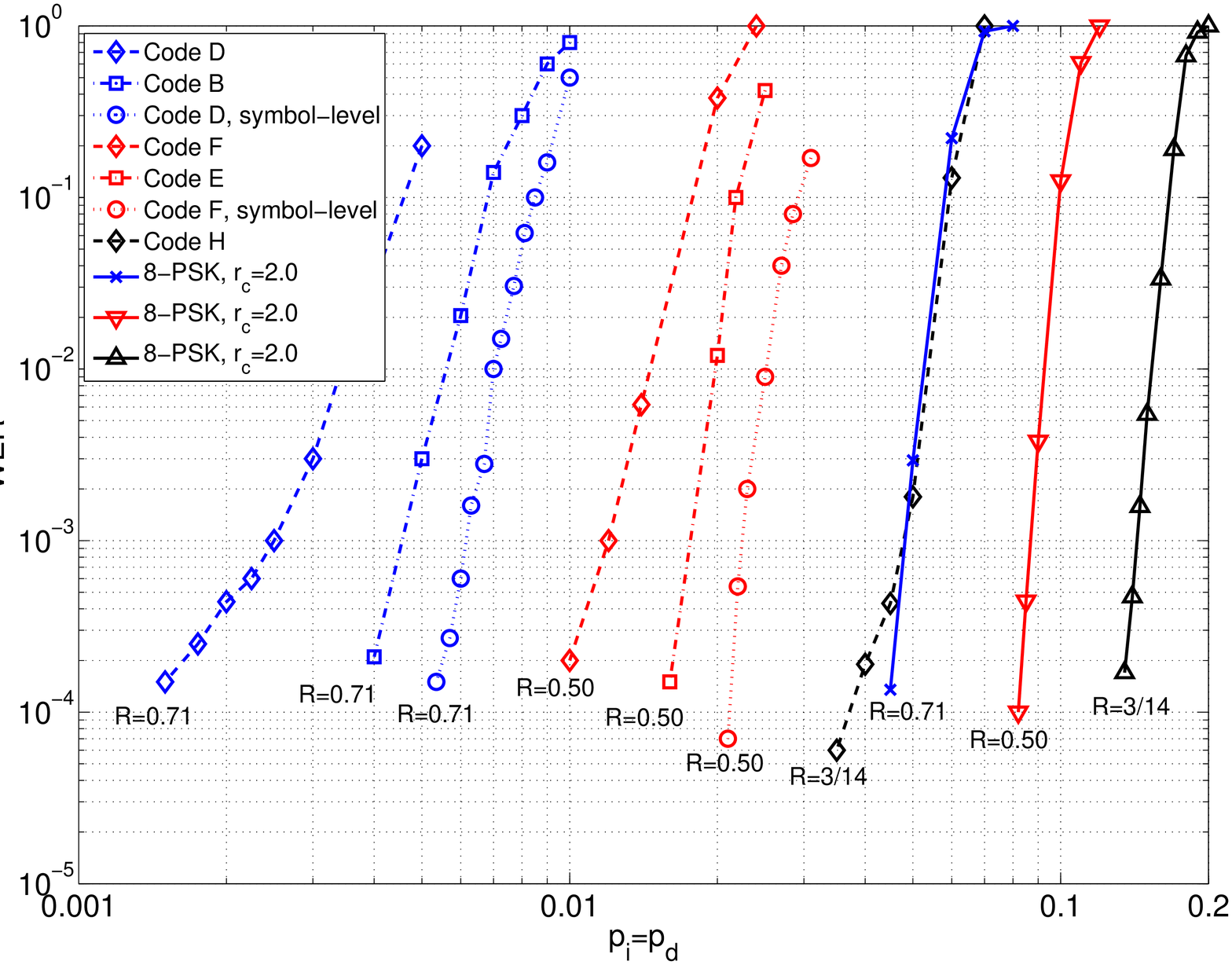}\\
\caption{WER comparison of the $8$-PSK watermarked system with the best results of \cite{davey01,ratzer05,briffa10}. Codes D, F, and H are binary watermark codes from \cite{davey01} with overall rates $0.71$, $0.50$, $3/14$, overall block lengths $4,995$, $4,002$, $4,662$, and outer LDPC codes defined over $\mathrm{GF(16)}$, $\mathrm{GF(16)}$, and $\mathrm{GF(8)}$, respectively. Code B and E are binary marker codes from \cite{ratzer05} with overall rates $0.71$ and $0.50$ and overall block lengths $4,995$ and $4,000$ with binary LDPC codes as outer codes. Codes D and F are also decoded by the symbol-level decoding method of \cite{briffa10}. All these codes are decoded on the binary I/D channel with no substitution errors or additive noise. For the non-binary I/D channel with $8$-PSK signalling in Example~1, we have done sliding window decoding at SNR$=\!20$~dB for three different LDPC codes with variable node degree $3$ and rates $0.71$, $0.50$, $3/14$ and block lengths $4,996$, $4,002$, $4,662$, respectively.}\label{fig:block_error_comparison}
\end{figure}

{\color{black}We are not aware of any practical coding method in the literature that can be directly and fairly compared to our proposed system. However, we provide comparisons with the best results of \cite{davey01,ratzer05,briffa10,wang11}. It is worth mentioning that this comparison is not completely fair as the I/D channel considered in these works is binary whereas in our case is non-binary. All in all, we believe that this comparison provides insight into what can be achieved by exploiting the extra degrees of freedom provided by non-binary signalling. This comparison is depicted in Fig.~\ref{fig:block_error_comparison}. To make the comparison as fair as possible, we have adjusted the block size and the rate of our $8$-PSK watermarked system according to the parameters of codes considered in the comparison. It is evident from Fig.~\ref{fig:block_error_comparison} that a significant improvement in the error correction performance is achieved by using non-binary signalling. There is also a significant performance improvement compared to the method of \cite{wang11} which considers marker codes with iterative exchange of information between the inner and outer decoders. Marker codes concatenated with optimized irregular LDPC outer codes with overall rates around $0.4$ and block length $5,000$ have been reported in \cite{wang11} which can reliably work under $\ppid<0.04$. As Fig.~\ref{fig:block_error_comparison} shows, a regular half-rate code with block length $4,002$ can do much better in our case even without iterative exchange of information.}

\subsection{Achievable information rates}
{\color{black}To obtain the capacity of the I/D channel, one is interested to calculate \cite{dobrushin67}
\begin{equation} \label{eq:ultimate_capacity}
C=\lim_{N\rightarrow\infty}\frac{1}{N}\sup_{P(\ubx)} I(\ubx;\uby),
\end{equation}
where $\ubx$ is the channel input sequence of length $N$, $\uby$ is the received sequence of random length, and $P(\ubx)$ denotes the joint distribution of the input sequence. Unfortunately, due to the presence of insertions/deletions, finding (\ref{eq:ultimate_capacity}) or its bounds has proven to be extremely challenging and the capacity is unknown. No single letter characterization of the mutual information also exists. Most of the results in the literature focus on sub-optimal decoding or more constrained channel models (such as deletion-only channel) and provide bounds on the capacity \cite{gallager61, zigangirov69, diggavi06, drinea07}. Most of these bounds, however, are driven for binary I/D channels and binary synchronization codes and either cannot be extended to non-binary I/D channel or become computationally extensive such as the bounds in \cite{fertonani11}.

A trellis-based approach is developed in \cite{hu10} to obtain achievable information rates for binary I/D channels with AWGN and inter-symbol interference under i.i.d. inputs (uniform $P(\ubx)$). This approach which mainly uses the forward pass of the forward-backward algorithm, can be extended to i.i.d. non-binary inputs and thus our channel model. We will use this method to find lower bounds on the capacity of the channel, i.e., $C_{\mathrm{i.u.d.}}$, and compare the achievable rates of our watermarked system with this lower bound. There also exist bounds on the performance of $q$-ary synchronization codes \cite{levenshtein02} which we will use in the comparisons.}

To obtain the achievable rates of our watermarked system, we calculate the maximum average per-symbol mutual information. In particular, we obtain an estimate of the average mutual information between $\uby$ and $\ubx$ given $\uw$. Assuming that $\ubx$ is a sequence of i.i.d. symbols, the average per-symbol mutual information is given by $\frac{1}{N}\sum_{i=1}^NI(\bx_i;\uby|\uw)$ where
\begin{align}
\nonumber I(\bx_i;\uby|\uw)=&\; H(\bx_i|w_i)-H(\bx_i|\uby,w_i)\\
=& \;r_\mathrm{c}-E_{\uby}\left[-\sum_{\bx_i\in\mathcal{X}^{w_i}}\left(\frac{P(\uby|\bx_i,w_i)}{\sum_{\bx_i\in\mathcal{X}^{w_i}}P(\uby|\bx_i,w_i)}
\log_2\left(\frac{P(\uby|\bx_i,w_i)}{\sum_{\bx_i\in\mathcal{X}^{w_i}}P(\uby|\bx_i,w_i)}\right)\right)\right],\label{eq:mutual_infor}
\end{align}
and the conditional probabilities are given by the watermark decoder. While it is not possible to do an exact calculation of the expectation, it is possible to calculate it numerically by Monte Carlo simulation. Then, (\ref{eq:mutual_infor}) can be used to find an estimate of the achievable rates of the watermarked system and a lower bound on the capacity of the channel. It should be noted that under large block lengths $N$, the variance of (\ref{eq:mutual_infor}) under Monte Carlo simulation is usually very small. Thus, it converges to the average very fast. Here, our results are averaged over $100$ blocks.

\begin{figure}
\centering
\includegraphics[width=.99\columnwidth]{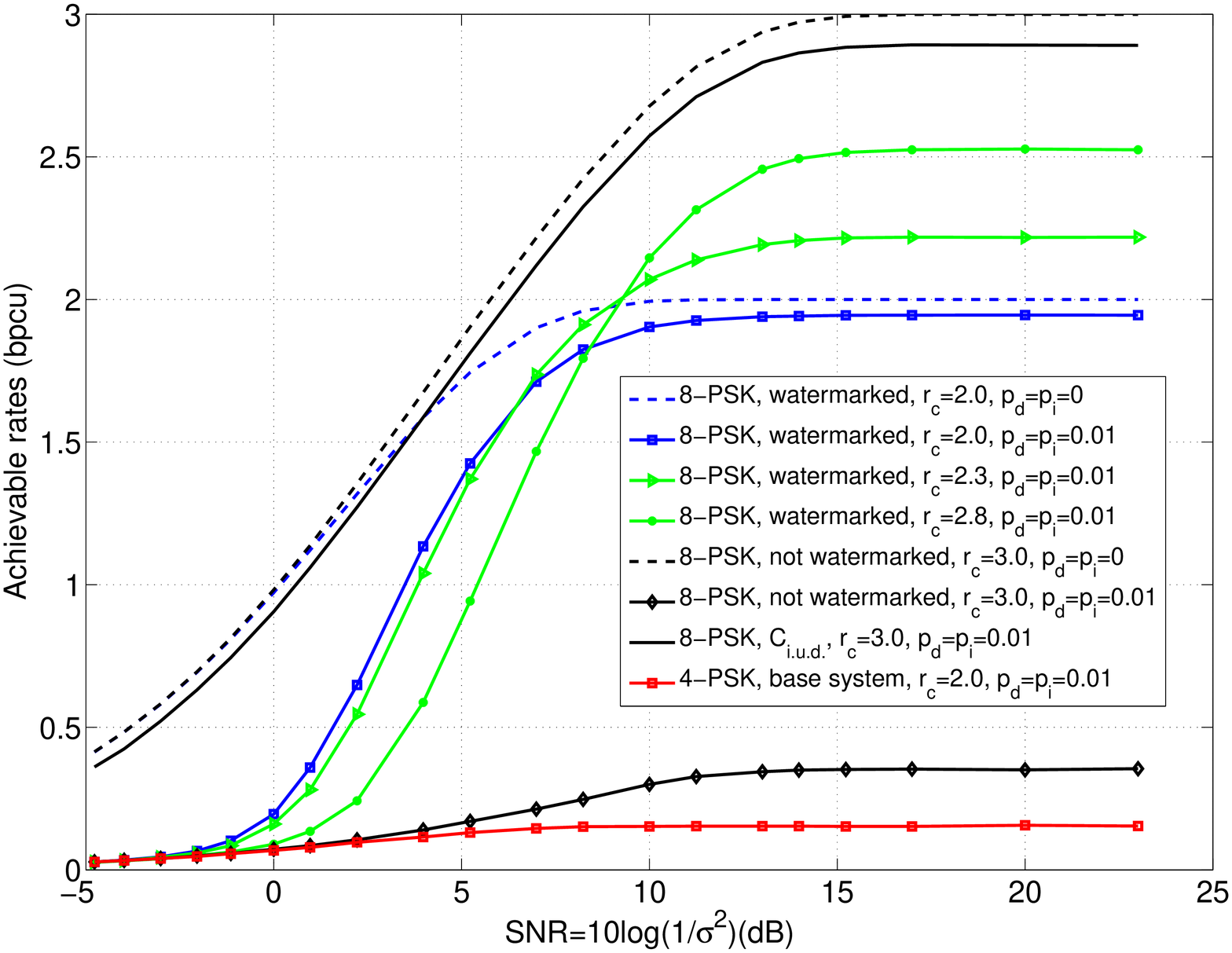}\\
\caption{Maximum achievable information rates (bits per channel use) versus SNR under different modulations assuming a $4$-PSK base system ($r_\mathrm{c}=2.0$). The $8$-PSK watermarked system mentioned in Section~\ref{Sec:modulator} has $r_\mathrm{c}=2.0$. The maximum achievable rates given by (\ref{eq:mutual_infor}) under $8$-PSK modulation with no watermark ($r_\mathrm{c}=3.0$), and under two $8$-PSK watermarked system with partial watermarking ($r_\mathrm{c}=2.3$ and $2.8$) mentioned in Section~\ref{sec:increasing_achievable_rates} are plotted for comparison. Also, $C_\mathrm{i.u.d.}$ is plotted for the $8$-PSK modulation.}\label{fig:entropy_8PSK}
\end{figure}

Using (\ref{eq:mutual_infor}) and assuming known block boundaries, the achievable information rates of the watermarked system of Example~1 is plotted versus SNR in Fig.~\ref{fig:entropy_8PSK}. We also compare the achievable rates with those of the two benchmark systems. One is the base system ($4$-PSK) which has the same $d_\mathrm{min}$, and another one is the system which has the same number of modulation points ($8$-PSK) as in the watermarked system but is not watermarked, i.e., all the $m+1$ bits are dedicated to information bits. Both of these systems use Gray mapping and are decoded by the forward-backward algorithm described in Section \ref{sec:watermark_decoder} with the exception that there is no watermark. The number of symbols per block is kept fixed at $10,012$ for all three systems so that the average number of symbols corrupted by insertions/deletions remains the same. {{\color{black}Notice that $r_\mathrm{c}=2.0$ for the base and the watermarked system and $r_\mathrm{c}=3.0$ for the $8$-PSK system with no watermark.}}

The dashed curves in Fig.~\ref{fig:entropy_8PSK} correspond to the maximum achievable information rates of the three systems when $\ppid=0$. In this case, it is clear that the watermarked and the base system achieve the same rates but the $8$-PSK system with no watermark achieves higher rates. At $\ppid=0.01$, however, the watermarked system performs much better than the two benchmark systems in terms of the maximum achievable rates (by using (\ref{eq:mutual_infor})) on the channel. This is of course not very surprising since no watermark is used in the benchmark systems and their only source of protection against insertions/deletions comes from the fact that they are decoded by the forward-backward algorithm.

{\color{black}The figure also depicts $C_\mathrm{i.u.d.}$ under $8$-PSK signalling at $\ppid=0.01$. Comparing this curve with the results given by (\ref{eq:mutual_infor}) shows how far the achievable rates of our watermarked system are from the maximum achievable rates on the channel using the same constellation under i.i.d. inputs. Although this gap is not small, we are not aware of any results in the literature that can approach $C_\mathrm{i.u.d.}$. This gap can be made smaller by the method we show in Section~\ref{sec:increasing_achievable_rates}.}

\begin{figure}
\centering
\includegraphics[width=.99\columnwidth]{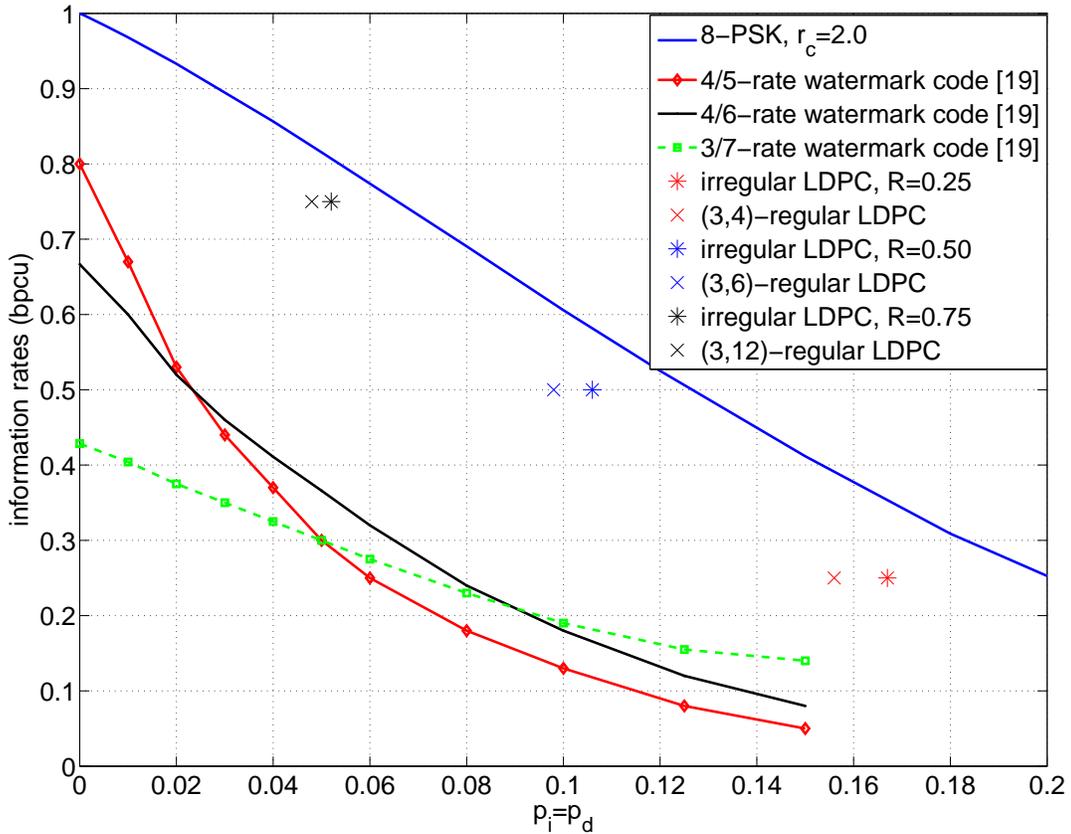}\\
\caption{Maximum achievable information rates as viewed by the outer code $I(d_{i,j};l_{i,j})$ for the $8$-PSK watermarked system at high SNR are compared with the obtained rates of \cite{davey01}. The rates of \cite{davey01} are calculated assuming no substitution errors. Also, the maximum $\ppid$ that the $8$-PSK watermarked system can tolerate with BER less than $10^{-5}$ is indicated for the three optimized irregular LDPC codes of rates $0.25$, $0.50$, and $0.75$ and three regular LDPC codes.}\label{fig:compare_with_davey}
\end{figure}

We also provide a comparison with \cite{davey01} in terms of comparing the achievable rates of these two systems as viewed by the outer code. In particular, we calculate the average $I(d_{i,j};l_{i,j})$ which is a number between $0$ and $1$ for both systems. This is done by Monte Carlo simulation, using the LLRs produced by the watermark decoder, i.e., using (\ref{eq:LLR}). This comparison is depicted in Fig.~\ref{fig:compare_with_davey}. The achievable rates seen by the outer code from \cite{davey01} are compared to those of the watermarked $8$-PSK system at SNR$=\!20$~dB. The rates of \cite{davey01} \textcolor{black}{are given for three binary watermark codes with sparsifier rates of $3/7$, $4/6$, and $4/5$ and} are calculated assuming no substitution error on the channel which is similar to the high SNR case on our channel. As depicted, the achievable rates of the proposed watermarked $8$-PSK system are much higher than those of \cite{davey01}.

Given the success of LDPC codes on many channels, we expect that the information rates of Fig.~\ref{fig:compare_with_davey} can be approached with carefully designed irregular LDPC codes of large block lengths. To demonstrate this, we have optimized the degree distributions of irregular LDPC codes of rates $0.25$, $0.50$, and $0.75$, and constructed codes of length $20,024$. The optimization process is done by the conventional numerical LDPC optimization methods in the literature (e.g., see \cite{chung01gaussian}). {\color{black} These optimization techniques usually use the pdf of the LLRs. On most channels, this LLR pdf can be calculated analytically. However, this cannot be done in our case due to nature of the channel. Thus, Monte Carlo simulation is used to find estimates of the LLR pdf. Given the channel parameters, this is done by simulating a large number of channel realizations, calculating LLRs using (\ref{eq:LLR}), and finally computing the average LLR pdf (probability mass function to be more precise). Next, the rate of the code is maximized by optimizing its degree distributions using the computed LLR pdf. The optimized degree distributions are given in Table.~\ref{Table}.} {\color{black}After optimizing the degree distributions, the parity-check matrices of the codes are constructed by the PEG algorithm \cite{Hu05}}. Finally, the codes are simulated on the channel at high SNR by assuming known block boundaries with the rest of parameters being the same as in Example~1. Fig.~\ref{fig:compare_with_davey} shows the maximum $\ppid$ under which the constructed irregular LDPC codes perform with BER less than $10^{-5}$. It is seen that these practically achievable rates are not far from the maximum achievable rates given by $I(d_{i,j};l_{i,j})$. Also depicted in Fig.~\ref{fig:compare_with_davey} are the results for three regular LDPC codes of the same length, i.e., $(3,4)$-regular, $(3,6)$-regular, and $(3,12)$-regular LDPC codes with rates $0.25$, $0.50$, and $0.75$, respectively.

\begin{table}[t]
\caption{Variable and check node degree distributions for the optimized irregular LDPC codes; All results achieved assuming maximum variable node degree of 30} \label{Table}
\begin{center}
\scriptsize
\begin{tabular}{|c|c|c|c|}
\hline & Code~1 & Code~2 & Code~3\\ \hline
$\lambda_2$ & 0.2793 & 0.1920 & 0.2562 \\
\hline
$\lambda_3$ & 0.2648 & 0.2480 & 0.3334\\
\hline
$\lambda_4$ &        & 0.0064 & 0.0010\\
\hline
$\lambda_5$ &        &        & 0.0022\\
\hline
$\lambda_6$ & 0.0173 & 0.0171 & 0.3621\\
\hline
$\lambda_7$ & 0.0575 & 0.0709 & 0.0434\\
\hline
$\lambda_8$ & 0.0938 & 0.1223 & 0.0017\\
\hline
$\lambda_9$ & 0.0279 & 0.0278 & \\
\hline
$\lambda_{10}$ & 0.0528 & 0.0302 & \\
\hline
$\lambda_{18}$ &        & 0.0413 & \\
\hline
$\lambda_{26}$ & 0.0494 & 0.0302 &\\
\hline
$\lambda_{27}$ & 0.0126 & 0.0137 &\\
\hline
$\lambda_{28}$ & 0.0179 & 0.0199 &\\
\hline
$\lambda_{29}$ & 0.0303 & 0.0358 &\\
\hline
$\lambda_{30}$ & 0.0964 & 0.1507 & \\
\hline
\hline $\rho_5$ & 1.0000&        & \\
\hline
$\rho_9$ &  & 1.0000 & \\
\hline
$\rho_{13}$ &  & & 1.0000\\
\hline
\hline
Rate & 0.25 & 0.50 & 0.75 \\
\hline
\end{tabular}
\end{center}
\end{table}

\begin{figure}
\centering
\includegraphics[width=.99\columnwidth]{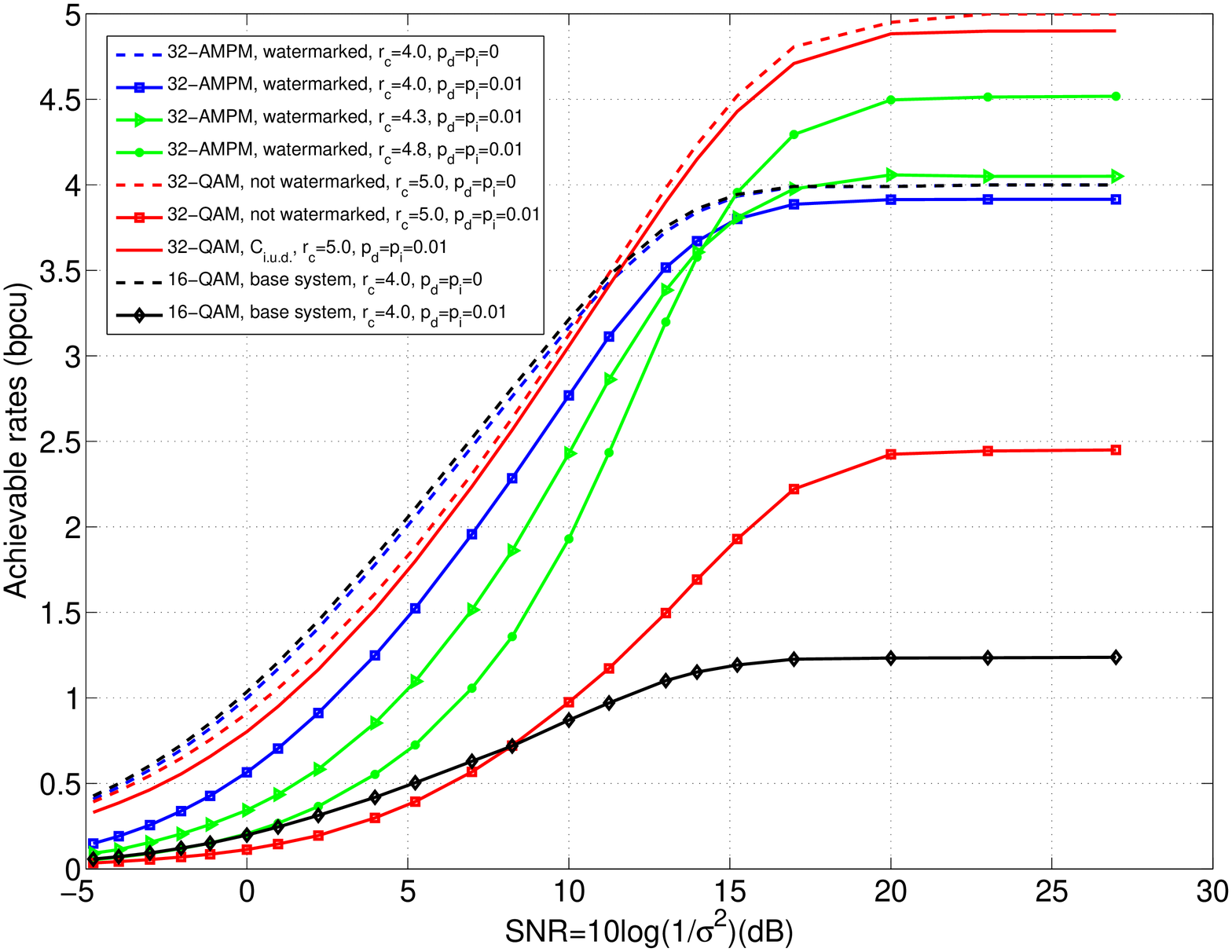}\\
\caption{Maximum achievable information rates (bits per channel use) versus SNR under different modulations assuming a $16$-QAM base system ($r_\mathrm{c}=4.0$). The $32$-AM/PM watermarked system mentioned in Section \ref{Sec:modulator} has $r_\mathrm{c}=4.0$. The maximum achievable rates given by (\ref{eq:mutual_infor}) under quasi-Gray $32$-QAM modulation with no watermark ($r_\mathrm{c}=5.0$), and under two $32$-AM/PM watermarked system with partial watermarking ($r_\mathrm{c}=4.3$ and $4.8$) mentioned in Section~\ref{sec:increasing_achievable_rates} are plotted for comparison. Also, $C_\mathrm{i.u.d.}$ is plotted for the $32$-QAM modulation.}\label{fig:entropy_32QAM}
\end{figure}

Now, consider Example~2. Using the same method, the maximum achievable information rates of the watermarked system (by using (\ref{eq:mutual_infor})) are compared to those of the two benchmark systems (Gray labeled $16$-QAM in Fig.~\ref{fig:16_QAM} and quasi Gray $32$-QAM) and $C_\mathrm{i.u.d.}$ in Fig.~\ref{fig:entropy_32QAM}. The block size is again kept fixed at $10,012$ symbols. The same discussion as in Example~1 applies to this case as well.

We also compare the maximum achievable information rates with the bounds available for $q$-ary synchronization codes. We use the asymptotic bounds for $q$-ary codes of \cite{levenshtein02} to compare with our scheme. {\color{black} These bounds are achieved by considering the Levenshtein distance \cite{levenshtein66} between $q$-ary codewords and enumerating the maximum size of codes capable of correcting insertions/deletions with \emph{zero} error probabilities}. Since these bounds do not consider substitution errors or additive noise, we compare them with the achievable rates of our scheme in the high SNR region.

\begin{figure}
\centering
\includegraphics[width=.99\columnwidth]{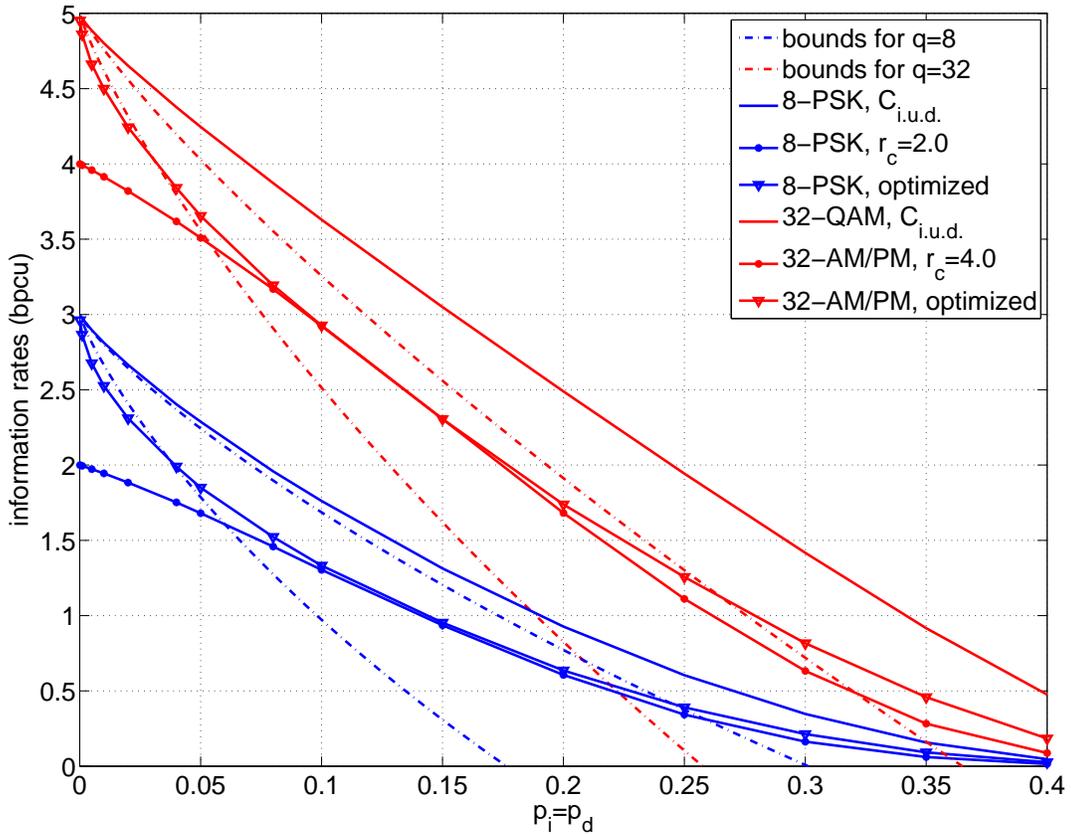}\\
\caption{Maximum achievable information rates of the watermarked systems at high SNR are compared to the upper and lower bounds of $q$-ary insertion/deletion correcting codes \cite{levenshtein02} and the achievable rates $C_\mathrm{i.u.d.}$. These rates are plotted for the $8$-PSK and $32$-AM/PM watermarked systems in two cases each. First, using binary watermark and assigning one bit to the watermark in each symbol ($r_\mathrm{c}=2.0$ and $r_\mathrm{c}=4.0$ for $8$-PSK and $32$-AM/PM, respectively). Second, the achievable rates are maximized by optimizing $q_\mathrm{w}$ and $r_\mathrm{c}$ in each point.}\label{fig:bounds}
\end{figure}

{\color{black}Fig.~\ref{fig:bounds} compares the upper and lower bounds of $q$-ary codes for $q=8,32$, with the achievable information rates of our $8$-PSK and $32$-AM/PM schemes using (\ref{eq:mutual_infor}), and also $C_\mathrm{i.u.d.}$. Notice that $C_\mathrm{i.u.d.}$ and our achievable information rates in some regions exceed the upper bound of \cite{levenshtein02}. This is due to the fact that the upper bounds are computed assuming zero error probabilities whereas $C_\mathrm{i.u.d.}$ and our achievable rates given by (\ref{eq:mutual_infor}) are computed assuming asymptotically vanishing error probabilities and thus computed in different scenarios. As seen on the figure, the achievable rates of our watermarked system is below $C_\mathrm{i.u.d.}$ and in some regions below the $q$-ary upper bound. Nevertheless, no code exists in the literature which can approach these limits. At small $\ppid$, the $q$-ary codes can theoretically achieve higher information rates than our scheme. This suggests that it is not efficient to dedicate one whole bit to the watermark when the number of synchronization errors is small. We will show in the next section how the information rates can be increased.}
\color{black}

\section{Increasing the achievable information rates}\label{sec:increasing_achievable_rates}
In this section, we show how the maximum achievable rates of the watermarked system can be improved.

{\color{black}We defined $r_\mathrm{c}$ to be the average number of bits assigned to the information bits (more precisely coded bits) per each transmitted symbol. Until now, we considered cases where one bit was assigned to the binary watermark in each of the transmitted symbols. For example, for the $4$-PSK base system and the $8$-PSK watermarked system discussed in Example~1, $r_\mathrm{c}=2.0$ bits. It is also possible to embed watermark bits into only some of the symbols but not all of them. For $8$-PSK, this means $2.0<r_\mathrm{c}<3.0$. We use Gray mapping for those symbols which are not watermarked. Also, we scatter those symbols which contain watermark uniformly in the transmitted block.}

It is evident that there is a trade-off between $r_\mathrm{c}$ and the system ability to recover synchronization errors. Increasing $r_\mathrm{c}$ potentially lets more information to pass through the channel but at the same time increases the system vulnerability to synchronization errors since less bits are assigned to the watermark. As a result, for a fixed signal set, there exists an optimum $r_\mathrm{c}$ for each $\ppid$ and SNR which provides the largest transmission rate on the channel.

For example consider the system of Example~1. At high SNR and $\ppid=0.01$, the maximum achievable rate is $1.945$ bits per channel use when $r_\mathrm{c}=2.0$ bits which increases to $2.528$ bits per channel use when $r_\mathrm{c}=2.8$ bits. This implies that dedicating one bit in every symbol to the watermark, i.e., $r_\mathrm{c}=2.0$ is wasteful in this case. A better protection is provided against synchronization errors by assigning one bit to the watermark in only $20\%$ of the symbols. As examples, Fig.~\ref{fig:entropy_8PSK} also shows the maximum achievable rates of the $8$-PSK system under $r_\mathrm{c}=2.3$ and $r_\mathrm{c}=2.8$. For $\ppid=0.01$, when SNR$<\!6.44$~dB the system with $r_\mathrm{c}=2.0$ achieves higher rates compared to those of systems with $r_\mathrm{c}=2.3$ and $r_\mathrm{c}=2.8$. When $6.44\!<$SNR$<\!9.31$, $r_\mathrm{c}=2.3$ provides the largest rates and when SNR$>\!9.31$~dB, $r_\mathrm{c}=2.8$ provides the largest rates compared to the other two systems. Fig.~\ref{fig:entropy_32QAM} depicts the comparison for the $16$-QAM and $32$-AM/PM systems.

Until now, we have only considered binary watermark sequences. When the number of synchronization errors is large, a binary watermark may not be very helpful in localizing these errors. Increasing the alphabet size of the watermark $q_\mathrm{w}$ increases the system ability to combat synchronization errors. Increasing $q_\mathrm{w}$, however decreases $r_\mathrm{c}$, i.e., less number of bits are available in each symbol for information bits. As a result, there is a trade-off between the $q_\mathrm{w}$, $r_\mathrm{c}$, and the maximum achievable information rate on the channel.

Fig.~\ref{fig:bounds} depicts the maximum achievable rates that $8$-PSK and $32$-AM/PM watermarked systems can achieve by finding the optimum $r_\mathrm{c}$ and $q_\mathrm{w}$ at each point. It is evident that the maximum achievable rates can be increased significantly by this strategy. This is especially beneficial when $\ppid$ is small where the achievable rates gets closer to the lower bound on $q$-ary codes.

\section{Complexity and the watermark sequence} \label{sec:complexity}
\subsection{Decoding complexity}
The complexity of the forward-backward algorithm determines the complexity of watermark decoding. This complexity scales as $O\left((1+2t_\mathrm{max})IMN\right)$, where $1+2t_\mathrm{max}$ is the number of states in the HMM and $N$ is equal to the number of symbols on which the forward-backward algorithm is performed. It should be noted that it is possible to reduce this complexity using arguments similar to those of \cite{davey01}.

%
\subsection{Watermark sequence}
The watermark sequences used in this paper are pseudo-random sequences. Our experiments confirm that these sequences perform well under different insertion/deletion rates. {\color{black}Periodic sequences with small periods are usually not good choices. First, they are vulnerable to successive insertios/deletions. As a simple example, if the watermark is a periodic sequence with period $4$, then the decoder cannot detect $4$ successive deletions. Furthermore, certain patterns of insertions/deletions can fool the decoder such that it fails to detect them. Randomness lowers the probability of false detecting or missing insertions/deletions by the decoder.

Among other factors which affect the decoding performance is the number of successive identical symbols (runs) in the watermark. One advantage of having runs is that it provides the ability to detect successive deletions. The larger the run-length, the larger successive deletions that can be detected. Nevertheless, larger runs lead to a worse localization of insertions/deletions as the decoder is not able to detect where exactly insertions/deletions have occurred. This gives rise to less reliable LLRs in the vicinity of insertions/deletions. Thus, there should be a balance between large and small runs in the watermark sequence. Pseudo-random sequences usually have this property.

Among candidates for the watermark are the run-length limited (RLL) sequences. RLL sequences with small maximum run-lengths (e.g., around 3 or 4) are good choices particularly when $\ppid$ is very small since the probability of having successive insertions/deletions is small. The performance gain over pseudo-random watermarks is only notable at small $\ppid$ where high-rate outer codes are used.

The presence of additive noise can also affect the choice of watermark. All in all, it is possible that sequences with structure could offer better performance than the above-mentioned sequences. This can be the subject of further investigation.}

\color{black}
\section{Conclusion} \label{Sec:conclusion}
In this paper, we proposed a concatenated coding scheme for reliable communication over non-binary I/D channels where symbols were randomly deleted from or inserted in the transmitted sequence and all symbols were corrupted by additive white Gaussian noise. First, we provided redundancy by expanding the symbol set while maintaining almost the same minimum distance. Then, we allocated part of the bits associated with each symbol to watermark bits. The watermark sequence, known at the receiver, was then used by the forward-backward algorithm to provide soft information for the outer code. Finally, the received sequence was decoded by the outer code.

We evaluated the performance of the watermarked system and through numerical examples we showed that significant amount of insertions and deletions could be corrected by the proposed method. The maximum information rates achievable by this method on the I/D channel were provided and compared with existing results and the available bounds on $q$-ary synchronization codes. Practical codes were also designed that could approach these information rates.

\bibliographystyle{IEEEtran}
\renewcommand{\baselinestretch}{1.5}
\bibliography{IEEEabrv,mybib}





\end{document}